\documentclass[useAMS,usenatbib]{mn2e}
\usepackage{graphicx}
\usepackage{amssymb}
\usepackage{color}

\def\be{\begin{equation}} 
\def\ee{\end{equation}}

\def\HI{\hbox{H~$\scriptstyle\rm I\ $}}

\def\gsim{\lower.5ex\hbox{\gtsima}} 
\def\lsim{\lower.5ex\hbox{\ltsima}} \def\gtsima{$\; \buildrel > \over 
\sim \;$} \def\ltsima{$\; \buildrel < \over \sim \;$} \def\prosima{$\; 
\buildrel \propto \over \sim \;$} \def\gsim{\lower.5ex\hbox{\gtsima}} 
\def\lsim{\lower.5ex\hbox{\ltsima}} 
\def\simgt{\lower.5ex\hbox{\gtsima}} 
\def\simlt{\lower.5ex\hbox{\ltsima}} 
\def\simpr{\lower.5ex\hbox{\prosima}} \def\la{\lsim} \def\ga{\gsim} 
  
 \def\gtsima{$\; \buildrel > \over \sim \;$} 
\def\ltsima{$\; \buildrel < \over \sim \;$} 
\def\gsim{\lower.5ex\hbox{\gtsima}} 
\def\lsim{\lower.5ex\hbox{\ltsima}} 
\def\simgt{\lower.5ex\hbox{\gtsima}} 
\def\simlt{\lower.5ex\hbox{\ltsima}} 
\def\simpr{\lower.5ex\hbox{\prosima}} 
\def\la{\lsim} 
\def\ga{\gsim}

\def\E3{{\cal E}_{\rm g}^{III}}

\def\Msun{\rm M_\odot}

\def\avchi{$\langle \chi_\mathrm{HI} \rangle$}
 
\title[21cm-LAE synergies]{Survey parameters for detecting 21cm - Ly$\alpha$ emitter cross correlations with the Square Kilometre Array}
\author[Hutter et al.]{Anne Hutter$^{1,2}$\thanks{E-mail: ahutter@swin.edu.au}, Cathryn M. Trott$^{2,3}$ \& Pratika Dayal$^{4}$\\ 
$^{{1}}$ Centre for Astrophysics \& Supercomputing, Swinburne University of Technology, Hawthorn, VIC 3122, Australia\\
$^{2}$ ARC Centre of Excellence for All Sky Astrophysics in 3 Dimensions (ASTRO 3D)\\
$^{3}$ International Centre for Radio Astronomy Research, Curtin University, Bentley WA 6102, Australia \\
$^{4}$ Kapteyn Astronomical Institute, University of Groningen, PO Box 800, 9700 AV Groningen, The Netherlands}
\begin{document} 
 
\date{} 
 
 
\maketitle 
 
\label{firstpage} 
\begin{abstract}
Detections of the cross correlation signal between the 21cm signal during reionization and high-redshift Lyman Alpha emitters (LAEs) are subject to observational uncertainties which mainly include systematics associated with radio interferometers and LAE selection. These uncertainties can be reduced by increasing the survey volume and/or the survey luminosity limit, i.e. the faintest detectable Lyman Alpha (Ly$\alpha$) luminosity. We use our model of high-redshift LAEs and the underlying reionization state to compute the uncertainties of the 21cm-LAE cross correlation function at $z\simeq6.6$ for observations with SKA1-Low and LAE surveys with $\Delta z=0.1$ for three different values of the average IGM ionization state (\avchi$\simeq0.1$, $0.25$, $0.5)$. 
At $z\simeq6.6$, we find SILVERRUSH type surveys, with a field of view of $21$~deg$^2$ and survey luminosity limits of $L_\alpha\geq7.9\times10^{42}$erg~s$^{-1}$, to be optimal to distinguish between an inter-galactic medium (IGM) that is $50$\%, $25$\% and $10$\% neutral, while surveys with smaller fields of view and lower survey luminosity limits, such as the $5$ and $10$~deg$^2$ surveys with WFIRST, can only discriminate between a $50$\% and $10$\% neutral IGM.
\end{abstract}

\section{Introduction}

The Epoch of Reionization marks the second major phase transition in the Universe, when ionizing photons from the first stars and galaxies gradually ionize the hydrogen in the intergalactic medium (IGM). Despite a number of observational constraints on the timing of reionization from quasar absorption lines \citep{Fan2006} and the cosmic microwave background \citep{Planck2016}, details of the progress, including reionization topology and the temporal and spatial evolution of the ionized regions, remain key open questions. 
On the one hand, detections of neutral hydrogen (\HI) through its 21cm emission using radio interferometers, including the Low Frequency Array (LoFAR), the Murchison Wide-field Array (MWA) and the forthcoming Square Kilometre Array (SKA), will be critical in shedding light on the propagation of ionized regions. On the other hand, the abundance and distribution of Lyman-$\alpha$ emitters (LAEs), galaxies identified by means of their Lyman-$\alpha$ (Ly$\alpha$) line at $1216$~\AA~ in the galaxy rest-frame, provide constraints on the mean \HI fraction \avchi~at $z\sim 5-8$ \citep[e.g.][]{dayal2008, dayal2011a, Jensen2013, Hutter2014}.

Given that the reionization state and topology will be hard to interpret from either dataset alone, recent efforts have focused on investigating the power of cross correlations between the 21cm signal and LAEs \citep{Wyithe2007, Vrbanec2016, Sobacchi2016, Hutter2017, Heneka2017, Wiersma2013}. Indeed, at a given \avchi~the amplitude of the 21cm-LAE cross correlation function on small scales is very similar for different reionization and LAE models \citep[cf.][]{Vrbanec2016, Sobacchi2016, Hutter2017, Kubota2017}. This is only because LAE galaxy identifications rely on sufficiently large ionized regions, either built up by themselves or neighbouring galaxies in clustered regions, and emitting enough Ly$\alpha$ photons into the IGM \citep{Castellano2016}. This implies that their positions are directly linked to the distribution of ionized regions and the overall ionization state of the IGM, making 21cm-LAE cross correlations a relatively robust measurement of \avchi~at a given epoch.

Low observational uncertainties will be critical in detecting the 21cm-LAE cross correlation signal and constraining \avchi. However, the reduction of the uncertainties arising from the 21cm signal measurements and the LAE observations favour opposite survey designs. While the uncertainties in the 21cm signal detection are reduced by larger survey volumes, the shot noise arising from the finite number of LAEs decreases with the survey limiting Ly$\alpha$ luminosity \citep{furlanetto-lidz2007, Kubota2017}. Sampling the Ly$\alpha$ luminosity function (Ly$\alpha$ LF), the number of LAEs rises quickly as the detectable Ly$\alpha$ luminosity is pushed to lower values.
These preferences lead to competing parameters for survey design, posing the question of which survey design (i.e. survey volume versus limiting Ly$\alpha$ luminosity) would be optimal and feasible to minimise the 21cm-LAE cross correlation uncertainties. In this paper, we address this question and compute the 21cm-LAE cross correlation uncertainties for various LAE Ly$\alpha$ luminosity limits and survey volumes by using the results of our numerical model for LAEs and reionization of the IGM at $z\simeq6.6$. 

The paper is organised as follows. In Section \ref{sec_model} we describe our numerical model for LAEs and reionization of the IGM at $z\simeq6.6$. We discuss the 21cm-LAE cross correlations for different survey depths in Section \ref{sec_crosscorr} and their associated observational uncertainties, for different survey strategies, in Section \ref{sec_observations}. We conclude in Section \ref{sec_conclusions}. Throughout this paper we assume a $\Lambda$CDM Universe with cosmological parameters values of $\Omega_\Lambda=0.73$, $\Omega_m=0.27$, $\Omega_b0.047$, $H_0=100h=70$km~s$^{-1}$Mpc$^{-1}$ and $\sigma_8=0.82$.

\begin{figure*}
 \includegraphics[width=0.78\textwidth]{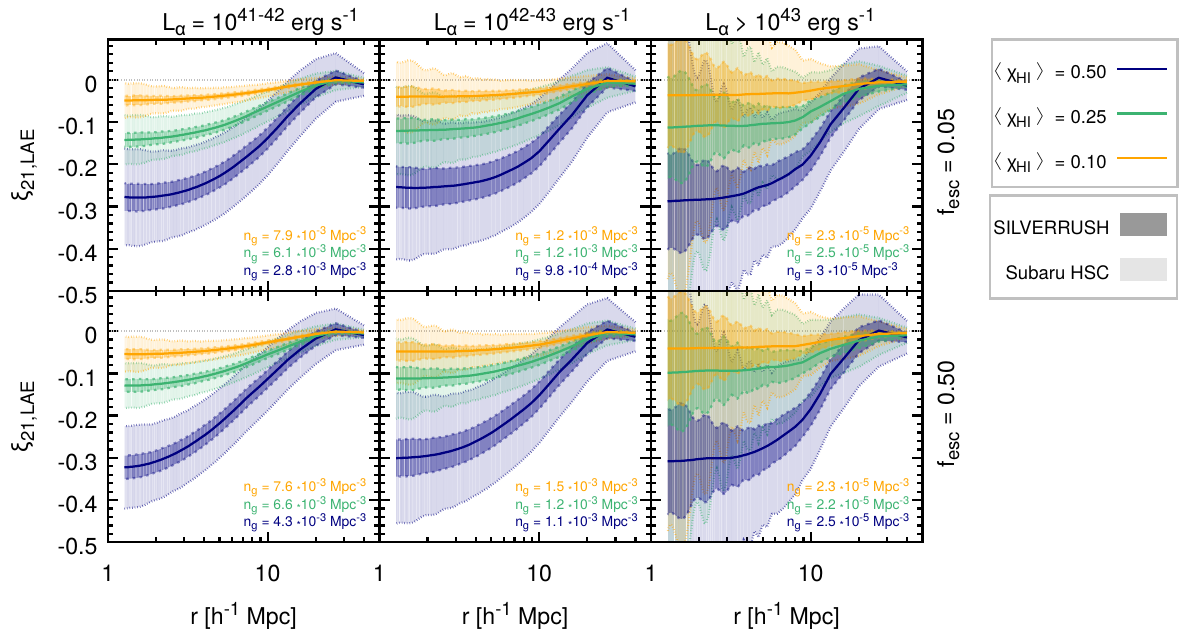}
 \caption{21cm-LAE cross correlation function for $f_\mathrm{esc}=0.05$ and $0.50$ (rows) and survey Ly$\alpha$ luminosity limits $L_\alpha=10^{41-42}$, $10^{42-43}$, $10^{>43}$erg~s$^{-1}$ (columns) at $z\simeq6.6$. Orange, green and blue lines represent the cross correlation functions at \avchi~$\simeq0.1$, $0.25$ and $0.5$, respectively. The light and dark shaded regions correspond to the values allowed by the uncertainties in computing the cross correlation between SKA and Subaru HSC or SILVERRUSH survey data. All identified LAEs have a minimum Ly$\alpha$ equivalent width, $EW_\alpha\geq20\AA$, and their corresponding number densities are indicated at the right bottom of each panel. The nearly constant amplitude across different Ly$\alpha$ luminosity limits shows that $\xi_\mathrm{21,LAE}$ is hardly sensitive to LAE clustering, which again increases with rising $L_\alpha$ values. However, stronger LAE clustering leads to rising uncertainties, as $P_\mathrm{LAE}$ in equation \ref{eq_uncertainties} increases.}
 \label{fig_crosscorr_21cm_LAE_subaru}
\end{figure*}

\begin{figure*}
 \includegraphics[width=0.76\textwidth]{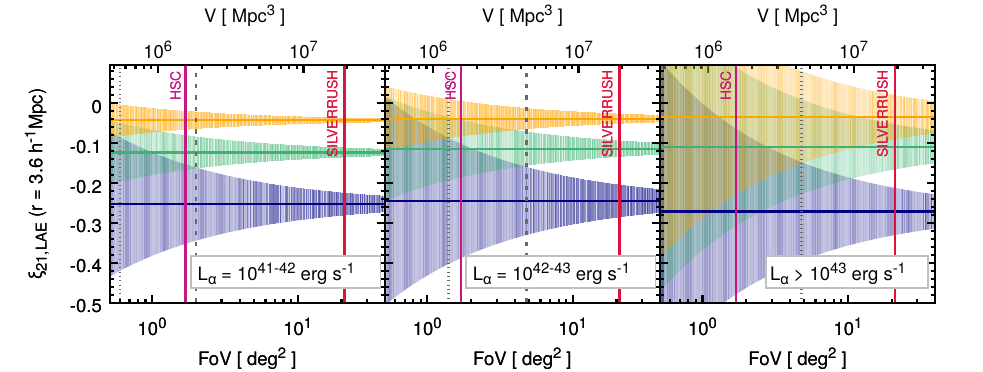}
 \caption{21cm-LAE cross correlation function at $r=3.6 h^{-1}$cMpc for $f_\mathrm{esc}=0.05$ and survey Ly$\alpha$ luminosity limits $L_\alpha=10^{41-42}$, $10^{42-43}$, $10^{>43}$erg~s$^{-1}$ at $z\simeq6.6$. Orange, green and blue lines represent \avchi~$\simeq0.1$, $0.25$ and $0.5$, respectively. The shaded regions show the cross correlation function uncertainties as a function of the survey volume of the SKA and LAE observations.}
 \label{fig_crosscorr_21cm_LAE_volume}
\end{figure*}

\section{Modelling LAEs \& the 21cm signal}
\label{sec_model}

Our model for $z\simeq6.6$ LAEs and the underlying reionization of the IGM combines a cosmological smoothed particle hydrodynamic (SPH) simulation run using {\sc gadget-2} with the {\sc pcrash} radiative transfer (RT) code and a model for ISM dust. We summarise the main characteristics of the model and refer the interested reader to \citet{Hutter2014} for detailed descriptions.

The hydrodynamical {\sc gadget-2} simulation has a box size of $80h^{-1}$~comoving Mpc (cMpc) and follows a total of $2\times1024^3$ dark matter (DM) and gas particles. It encompasses physical descriptions for star formation, metal production and feedback as described in \citet{springel2003}, and assumes a \cite{salpeter1955} initial mass function (IMF) between $0.1-100\Msun$. In our analysis, we consider only ``resolved'' galaxies within the simulation that contain at least $10$ star particles and halo masses $M_h>10^{9.2}\Msun$. For each galaxy the intrinsic spectrum is derived by summing over all the spectra of its star particles using with the stellar population synthesis code {\sc starburst99} \citep{leitherer1999}. The dust mass produced by Type II SN (SNII) during the first billion years and the corresponding attenuation of ultra-violet (UV) radiation are computed following the dust model described in \citet{dayal2010}. The observed UV luminosity can be calculated as $L_c^\mathrm{obs}=f_c\times L_c^\mathrm{int}$, where $L_c^\mathrm{int}$ is the intrinsic UV luminosity and $f_c$ the fraction of UV photons that escape the ISM unattenuated by dust.
The observed Ly$\alpha$ luminosity is computed as $L_{\alpha}^\mathrm{obs}=L_{\alpha}^\mathrm{int} f_{\alpha} T_{\alpha}$ where $f_\alpha$ and $T_{\alpha}$ account for the Ly$\alpha$ attenuation by ISM dust and IGM \HI, respectively. Galaxies with a Ly$\alpha$ equivalent width $EW_\alpha=L_\alpha^\mathrm{obs}/L_c^\mathrm{obs} \geq 20$\,\AA\ and a chosen $L_\alpha$ lower luminosity limit are identified as LAEs.
In order to derive $T_\alpha$ for each galaxy at different \avchi~values, the $z\simeq6.6$ snapshot of the hydrodynamical simulation is post-processed with the RT code {\sc pcrash}. For $5$ different values for the escape fraction of ionizing photons from the galaxies, $f_\mathrm{esc}=0.05$, $0.25$, $0.5$, $0.75$, $0.95$, {\sc pcrash} computes the evolution of the ionized regions resulting from the ionizing radiation of $\sim 3\times10^5$ ``resolved'' galaxies, and is run until the IGM is fully ionized. In order to fit our LAE model to the observed Ly$\alpha$ LF at $z\simeq6.6$ \citep{Kashikawa2011}, the only free parameter is the ratio between the escape fractions of Ly$\alpha$ and UV continuum photons, $p=f_{\alpha}/f_c$ (for values see Table 1 in \citet{Hutter2014}).
For all allowed parameter combinations of $f_\mathrm{esc}$, \avchi~and $p$, we derive the differential 21cm brightness temperature fields from the respective ionization field following \citet{iliev2012}.
\begin{eqnarray}
\delta T_b (\vec{x}) &=& T_0\ \langle \chi_\mathrm{HI} \rangle \ \left[1+\delta(\vec{x})\right] \ \left[1+\delta_\mathrm{HI}(\vec{x})\right] \\
T_0&=&28.5 \mathrm{mK}\ \left(\frac{1+z}{10}\right)^{1/2} \frac{\Omega_b}{0.042} \frac{h}{0.073} \left(\frac{\Omega_m}{0.24}\right)^{-1/2}
\label{eq_Tb}
\end{eqnarray}
Here, $1+\delta(\vec{x})=\rho(\vec{x})/\langle\rho\rangle$ and $1+\delta_\mathrm{HI}(\vec{x})=\chi_\mathrm{HI}(\vec{x})/\langle \chi_\mathrm{HI}\rangle$ refer to the local gas density and \HI fraction compared to their corresponding average global values, respectively. 

\section{21cm-LAE cross correlations}
\label{sec_crosscorr}

In order to determine the best survey design to constrain the neutral hydrogen fraction of the IGM during reionization, we compute the cross correlation functions between the 21cm signal and $z\simeq6.6$ LAEs using $3$ luminosity cuts in $L_\alpha=10^{41-42}$ (faint LAEs; LAE$_\mathrm{f}$), $10^{42-43}$ (intermediate LAEs; LAE$_\mathrm{i}$) and $10^{>43}$ erg~s$^{-1}$ (bright LAEs; LAE$_\mathrm{b}$). We derive the dimensionless cross correlation functions for each limiting luminosity as
\begin{eqnarray}
 \xi_\mathrm{21,LAE}(r) &=& \int P_\mathrm{21,LAE}(k)\ \frac{\sin(kr)}{kr}\ 4\pi k^2\ \mathrm{d}k.
 \label{eq_crosscorr}
\end{eqnarray}
Here the cross power spectrum $P_\mathrm{21,LAE}(k)=V \langle \tilde{\Delta}_\mathrm{21} ({\bf k})\ \tilde{\Delta}_\mathrm{LAE}(-\bf{k}) \rangle$ is in units of Mpc$^3$  and derived from the product of the Fourier transformation\footnote{The Fourier transformation of $\Delta({\bf x})$ is computed as $\tilde{\Delta}({\bf k}) = V^{-1} \int \Delta({\bf x})\ e^{-2\pi i{\bf k x}}\ \mathrm{d}^3x$.} of the fractional fluctuation fields of the 21cm signal, $\delta_\mathrm{21} = \delta T_b/T_0$, and the LAE number density, $\delta_\mathrm{LAE} = n_\mathrm{LAE}/\langle n_\mathrm{LAE}\rangle -1$.

In Fig. \ref{fig_crosscorr_21cm_LAE_subaru} the solid lines show $\xi_\mathrm{21,LAE}$  at various stages of reionization (\avchi$\simeq0.5$, $0.25$, $0.1$) for two different ionizing escape fractions, $f_\mathrm{esc}=0.05$, $0.5$. We note that parameter combinations used in this work are consistent with the LAE Ly$\alpha$ LF at $z=6.6$. As expected $\xi_\mathrm{21,LAE}$ indicates an anti-correlation between the 21cm signal and LAEs on scales smaller than the average size of the ionized regions around LAEs. 
With the IGM becoming more ionized, the abundance of LAEs increases and the mean 21cm differential brightness temperature, $\langle \delta T_b\rangle$, drops. The latter decreases the contrast between $\delta T_b$ at LAE locations and $\langle \delta T_b\rangle$, leading to a weaker anti-correlation. However, the anti-correlation strength also depends on the residual \HI fraction within the ionized regions around LAEs \citep{Hutter2017}. With decreasing $f_\mathrm{esc}$, the photoionization rate ($\Gamma_\mathrm{HI}$) drops and the residual \HI fraction increases, which causes a slightly weaker anti-correlation for $f_\mathrm{esc}=0.05$ than for $0.5$. The lower ionization fractions in ionized regions are compensated by slightly larger ionized regions, which become apparent in the anti-correlation extending to larger scales.

The extent and strength of the anti-correlation between the 21cm signal and LAEs reflect the size and the degree of ionization of the ionized regions around the selected LAEs, respectively. With L$_\alpha$ being directly proportional to the number of ionizing photons produced in a galaxy, the sizes of the ionized regions around LAEs rise from faint to bright LAEs, e.g. for $f_\mathrm{esc}=0.5$ and \avchi$\simeq0.5$, $\xi_\mathrm{21,LAE}$ drops from $-0.23$ for LAE$_\mathrm{f}$ to $-0.3$ for LAE$_\mathrm{b}$ at $r=5h^{-1}$cMpc. 
Comparing the anti-correlation strengths across the L$_\alpha$ bins, we notice the strength to increase towards fainter LAEs for a mostly ionized IGM (\avchi$<0.3$): fainter LAEs are more likely to be located in less over-dense regions, leading to lower residual \HI fractions in their ionized regions.
In contrast, for \avchi$\simeq0.5$, the anti-correlation strength is stronger for LAE$_\mathrm{b}$ than for LAE$_\mathrm{i}$. At these earlier stages of reionization, the equilibrium \HI fraction in the ionized regions has not been reached, thus the photoionization rate and  ionization fraction close to the brightest galaxies are the highest. Furthermore, in contrast to LAE$_\mathrm{i}$, LAE$_\mathrm{f}$ are only found in clustered regions around bright galaxies that provide enough ionizing emissivity to keep the region ionized.

\section{Observational uncertainties}
\label{sec_observations}

We derive the observational uncertainties of the 21cm-LAE cross correlations from the cross power spectra uncertainties, which include sample variance ($P_\mathrm{21}$) and thermal noise ($\sigma_\mathrm{21}$) from the 21cm signal as well as sample variance ($P_\mathrm{LAE}$) and shot noise ($\sigma_\mathrm{LAE}$) from LAEs as
\begin{eqnarray}
 \delta P_\mathrm{21,LAE}^2({\bf k}) &=& 2\ P_\mathrm{21,LAE}^2({\bf k})  \label{eq_uncertainties} \\
 &+& 2\ \left[P_\mathrm{21}({\bf k}) + \sigma_\mathrm{21}^2({\bf k})\right]\left[P_\mathrm{LAE}({\bf k}) + \sigma_\mathrm{LAE}^2({\bf k})\right]. \nonumber
\end{eqnarray}
The thermal noise depends on the characteristics of the radio interferometer, $\sigma_\mathrm{21}^2 ({\bf k})= \frac{T_\mathrm{sys}^2/ T_0^2}{N_b({\bf k})\ \Delta\nu\ \Delta t}\frac{V}{(2\pi)^3}$. This includes its system temperature ($T_\mathrm{sys}$), the number of baselines contributing to angular mode $(k_x,k_y)$ ($N_b$), its band width ($\Delta \nu$), and the observed volume ($V$) and integration time ($\Delta t$). The shot noise arising from the finite number of LAEs is determined by their mean number density $n_\mathrm{LAE}$, $\sigma_\mathrm{LAE}^2 ({\bf k}) = \left[(2\pi)^3 n_\mathrm{LAE}\right]^{-1}$.
In a next step, we compute the spherically averaged cross power spectra uncertainties $\delta P_\mathrm{21,LAE}^2(k) = \delta P_\mathrm{21,LAE}^2({\bf k}) / N(k)$, where $N(k)$ denotes the number of modes in each $k = \sqrt{k_x^2+k_y^2+k_z^2}$ bin. Uncertainties of the cross correlation functions are derived by propagating the cross power spectra uncertainties following equation \ref{eq_crosscorr}, while assuming that different $k$ bins are correlated. The level of independence between $k$ bins is determined by the SKA1-Low station size, and the array baseline layout.

To determine the best survey design for detecting $\xi_\mathrm{21,LAE}$ with SKA1-Low, we assume an integration time of $1000$h and the array configuration V4A\footnote{\texttt{http://astronomers.skatelescope.org/wp-content/uploads/\\*2015/11/SKA1-Low-Configuration\_V4a.pdf}}. The latter results in a filling factor that reduces substantially outside the core, yielding poorer brightness temperature sensitivity performance on small scales. Temperature and effective collecting area as a function of frequency are matched to the systemic specification in {\it SKA1 System Baseline Design} document\footnote{\texttt{http://astronomers.skatelescope.org/wp-content/uploads/\\*2016/05/SKA-TEL-SKO-0000002\_03\_SKA1SystemBaselineDesignV2.pdf}}.

We derive the cross correlation uncertainties ($\delta \xi_\mathrm{21,LAE}$) at $z\simeq6.6$ directly from our $80h^{-1}$cMpc simulation box except for the survey volume, which we treat as a free parameter. We consider a survey at $z\simeq6.6$ with a line-of-sight depth corresponding to $\Delta z=0.1$ and various field of views (FoV) that are within the SKA FoV limits. We note that feasible LAE surveys are generally smaller in volume than the 21cm surveys with SKA.

The bright and dark shaded regions in Fig. \ref{fig_crosscorr_21cm_LAE_subaru} show the 21cm-LAE cross correlation uncertainties, $\delta \xi_\mathrm{21,LAE}$, for a survey area of $1.8$ and $21$~deg$^2$, respectively, corresponding to the FoVs of Hyper Suprime-Cam (HSC) on Subaru Telescope and the SILVERRUSH survey \citep{Ouchi2018}. As expected, $\delta \xi_\mathrm{21,LAE}$ decreases as the survey volume increases (HSC vs. SILVERRUSH) and as the number density of LAEs, $n_\mathrm{LAE}$, rises towards fainter Ly$\alpha$ luminosities. The signal-to-noise-ratio (SNR) varies with spatial scale $r$. It drops rapidly as soon as scales $r$ exceed the average size of the ionized regions around LAEs ($R_\mathrm{ion}$), caused by the decline in the anti-correlation amplitude. With the anti-correlation being strongest on scales $r<R_\mathrm{ion}$, the SNR is highest on small scales, with the optimal scale increasing with the Ly$\alpha$ luminosity limit. An increasing Ly$\alpha$ luminosity limit corresponds to a decreasing LAE number density and thus poorer sensitivity to variations on smaller and smaller scales. This decline in sensitivity leads to a drop in the SNR on small scales, visible for LAE$_\mathrm{b}$ at $r\lesssim4h^{-1}$cMpc. Hence, the best SNR values are obtained at intermediate scales. 
Thus, we show the $\delta \xi_\mathrm{21,LAE}$ values at $r=3.6h^{-1}$cMpc as a function of the survey volume in Fig. \ref{fig_crosscorr_21cm_LAE_volume}, which allow us to identify the minimum survey volume to distinguish between \avchi$\simeq0.1$, $0.25$ and $0.5$ (\avchi$\simeq0.1$ and $0.5$). Assuming that overlapping shaded regions do not allow a differentiation between the respective ionization states, we obtain the minimum FoVs required for detection, indicated by the long-dashed (dashed) gray vertical lines: $2.0$, $4.8$, $48$~deg$^2$ ($0.6$, $1.4$, $4.8$~deg$^2$) for $L_\alpha=10^{41-42}$, $10^{42-43}$, $10^{>43}$erg~s$^{-1}$. We note that the FoV required for LAE$_\mathrm{b}$ exceeds the SKA FoV of $37$~deg$^2$.

\begin{figure}
 \includegraphics[width=0.47\textwidth]{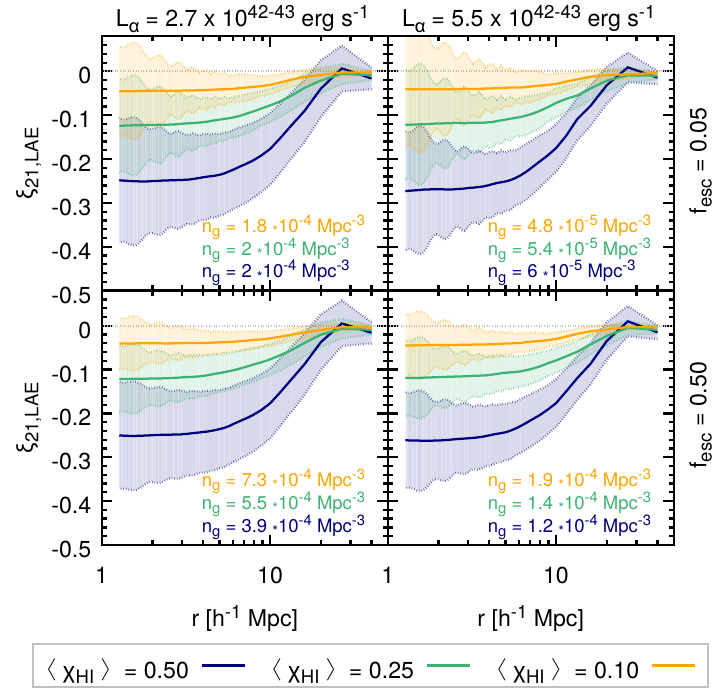}
 \caption{Same as Fig. \ref{fig_crosscorr_21cm_LAE_subaru} but for $L_\alpha=2.7\times10^{41-42}$, $5.5\times10^{42-43}$erg~s$^{-1}$, equivalent to WFIRST survey luminosity limits.}
 \label{fig_crosscorr_21cm_LAE_wfirst}
\end{figure}

From Fig. \ref{fig_crosscorr_21cm_LAE_volume} we see that HSC can only distinguish between \avchi$\simeq0.1$ and $0.5$ for $L_\alpha<10^{43}$erg~s$^{-1}$, while the $\sim12$ times larger FoV of the SILVERRUSH survey allows this differentiation for LAE$_\mathrm{b}$. SILVERRUSH FoVs in combination with LAE$_\mathrm{i}$ are even sufficient to distinguish between \avchi$\simeq0.1$, $0.25$ and $0.5$.
Finally, we show the 21cm-LAE correlation functions and their uncertainties for $5$ and $10$~deg$^2$ surveys planned with WFIRST in Fig. \ref{fig_crosscorr_21cm_LAE_wfirst}, with limiting Ly$\alpha$ luminosities of $2.7\times10^{42}$ and $5.5\times10^{42}$erg~s$^{-1}$, respectively. Here the scale dependence of the SNR is key, as the $5$~deg$^2$ FoV survey can only distinguish between \avchi$\simeq0.1$ and $0.5$ on scales of $r>2h^{-1}$cMpc, and the $10$~deg$^2$ FoV survey between \avchi$\simeq0.1$, $0.25$ and $0.5$ on scales of $r=5-10h^{-1}$cMpc.

\section{Conclusions}
\label{sec_conclusions}

In this letter, we explore the best suited and feasible survey designs to detect the cross correlation between the 21cm signal and LAEs at $z\simeq6.6$ with SKA1-Low. From our reionization simulations, we compute the 21cm-LAE cross correlations at \avchi$=(0.1$, $0.25$, $0.5)$ for multiple Ly$\alpha$ luminosity bins (faint, intermediate, bright) corresponding to different survey luminosity limits. Following the extent of the ionized regions around LAEs, the anti-correlation extends to increasingly larger scales as brighter LAEs are considered, while its strength is only marginally affected, indicating that cross correlations are hardly sensitive to LAE clustering. 

We briefly note that this parameter space is much larger than the ($3-\sigma$) constraints of \avchi $\lsim 0.01$ derived, using the {\it mean} LAE angular correlation function (ACF), averaged over multiple sub-volumes and lines of sight, in \citet{Hutter2015}. However, given the patchiness of reionization and the line of sight dependence of Ly$\alpha$ transmission, the lower limit of the ACFs \citep[Fig. 1;][]{Hutter2015} are consistent with \avchi$ = 0.1, 0.25$ at all scales and with \avchi$ =0.5$ (except at the very smallest scales). Given the power of 21cm-LAE cross correlations in determining the history and topology of reionization, in this work, we explore a much larger parameter space.

For all cross correlations we derive the corresponding observational uncertainties from 21cm measurements with SKA1-Low and an arbitrary high-redshift LAE survey with $\Delta z=0.1$. Given that these uncertainties decrease with larger survey volumes and lower survey limiting Ly$\alpha$ luminosities, we find that for a survey limiting luminosity $L_\alpha>10^{42}$erg~s$^{-1}$ a survey field of view of at least $5$~deg$^2$ is needed. Lower survey limiting Ly$\alpha$ luminosities require larger survey volumes, however, around $L_\alpha\sim10^{43}$erg~s$^{-1}$, LAE number densities become so low that the mitigation of the associated shot noise requires field of views exceeding that of SKA.
LAE surveys with large field of views and detecting the intermediate to bright LAEs, such as SILVERRUSH with $21$~deg$^2$ and $L_\alpha\geq7.9\times10^{42}$erg~s$^{-1}$ at $z\simeq6.6$ \citep{Ouchi2018}, are optimal to distinguish between an IGM that is $10$\%, $25$\% and $50$\% neutral. $5$ and $10$~deg$^2$ survey with WFIRST allow a distinction between \avchi$\simeq0.1$ and $0.5$ at intermediate scales ($r\simeq3-10 h^{-1}$cMpc). 

Certainly, observational uncertainties increase with stronger LAE clustering as long as they are not dominated by the LAE shot noise, as in e.g. the SILVERRUSH survey. Our simulated $z\simeq6.6$ LAEs, however, are rather more than less clustered than the observed ones.\footnote{We computed $\delta\xi_\mathrm{21,LAE}$ for a reduced LAE clustering, where we used $0.5$ times the LAE power spectra for \avchi$=10^{-4}$, which is in excellent agreement with observations \citep{Hutter2015}. While $\delta\xi_\mathrm{21,LAE}$ does not change for LAE$_\mathrm{b}$, it drops marginally (by a factor 2) for \avchi$=0.25$, $0.1$ ($0.5$) for LAE$_\mathrm{i}$ and LAE$_\mathrm{f}$.} Nevertheless, as LAE number densities and clustering are $z$-dependent, the $z$-evolution of the 21cm-LAE cross correlation uncertainties may alter optimal survey parameters and further studies are required to determine the best survey designs at higher-$z$.

\section*{Acknowledgements} 
AH is supported by the Australian Research Council's Discovery Project funding scheme (DP150102987). Parts of this research were supported by the Australian Research Council Centre of Excellence for All Sky Astrophysics in 3 Dimensions (ASTRO 3D; CE170100013). PD acknowledges support from the European Research Council's starting grant ERC StG-717001 ``DELPHI" and from the CO-FUND Rosalind Franklin program. We acknowledge support from the Munich Institute for Astro- and Particle Physics of the DFG cluster of excellence ``Origin and Structure of the Universe".

\bibliographystyle{mn2e}
\bibliography{synergy}

\begin{thebibliography}{23}
\expandafter\ifx\csname natexlab\endcsname\relax\def\natexlab#1{#1}\fi

\bibitem[{{Castellano} {et~al}\mbox{.}(2016){Castellano}, {Dayal},
  {Pentericci}, {Fontana}, {Hutter}, {Brammer}, {Merlin}, {Grazian}, {Pilo},
  {Amorin}, {Cristiani}, {Dickinson}, {Ferrara}, {Gallerani}, {Giallongo},
  {Giavalisco}, {Guaita}, {Koekemoer}, {Maiolino}, {Paris}, {Santini},
  {Vallini}, {Vanzella}, \& {Wagg}}]{Castellano2016}
{Castellano} M. {et~al.}, 2016, \apjl, 818, L3

\bibitem[{{Dayal} {et~al}\mbox{.}(2008){Dayal}, {Ferrara}, \&
  {Gallerani}}]{dayal2008}
{Dayal} P., {Ferrara} A., {Gallerani} S., 2008, \mnras, 389, 1683

\bibitem[{{Dayal} {et~al}\mbox{.}(2010){Dayal}, {Ferrara}, \&
  {Saro}}]{dayal2010}
{Dayal} P., {Ferrara} A., {Saro} A., 2010, \mnras, 402, 1449

\bibitem[{{Dayal} {et~al}\mbox{.}(2011){Dayal}, {Maselli}, \&
  {Ferrara}}]{dayal2011a}
{Dayal} P., {Maselli} A., {Ferrara} A., 2011, \mnras, 410, 830

\bibitem[{{Fan} {et~al}\mbox{.}(2006){Fan}, {Carilli}, \& {Keating}}]{Fan2006}
{Fan} X., {Carilli} C.~L., {Keating} B., 2006, \araa, 44, 415

\bibitem[{{Furlanetto} \& {Lidz}(2007)}]{furlanetto-lidz2007}
{Furlanetto} S.~R., {Lidz} A., 2007, \apj, 660, 1030

\bibitem[{{Heneka} {et~al}\mbox{.}(2017){Heneka}, {Cooray}, \&
  {Feng}}]{Heneka2017}
{Heneka} C., {Cooray} A., {Feng} C., 2017, \apj, 848, 52

\bibitem[{{Hutter} {et~al}\mbox{.}(2015){Hutter}, {Dayal}, \&
  {M{\"u}ller}}]{Hutter2015}
{Hutter} A., {Dayal} P., {M{\"u}ller} V., 2015, \mnras, 450, 4025

\bibitem[{{Hutter} {et~al}\mbox{.}(2017){Hutter}, {Dayal}, {M{\"u}ller}, \&
  {Trott}}]{Hutter2017}
{Hutter} A., {Dayal} P., {M{\"u}ller} V., {Trott} C.~M., 2017, \apj, 836, 176

\bibitem[{{Hutter} {et~al}\mbox{.}(2014){Hutter}, {Dayal}, {Partl}, \&
  {M{\"u}ller}}]{Hutter2014}
{Hutter} A., {Dayal} P., {Partl} A.~M., {M{\"u}ller} V., 2014, \mnras, 441,
  2861

\bibitem[{{Iliev} {et~al}\mbox{.}(2012){Iliev}, {Mellema}, {Shapiro}, {Pen},
  {Mao}, {Koda}, \& {Ahn}}]{iliev2012}
{Iliev} I.~T., {Mellema} G., {Shapiro} P.~R., {Pen} U.-L., {Mao} Y., {Koda} J.,
  {Ahn} K., 2012, \mnras, 423, 2222

\bibitem[{{Jensen} {et~al}\mbox{.}(2013){Jensen}, {Laursen}, {Mellema},
  {Iliev}, {Sommer-Larsen}, \& {Shapiro}}]{Jensen2013}
{Jensen} H., {Laursen} P., {Mellema} G., {Iliev} I.~T., {Sommer-Larsen} J.,
  {Shapiro} P.~R., 2013, \mnras, 428, 1366

\bibitem[{{Kashikawa} {et~al}\mbox{.}(2011){Kashikawa}, {Shimasaku}, {Matsuda},
  {Egami}, {Jiang}, {Nagao}, {Ouchi}, {Malkan}, {Hattori}, {Ota}, {Taniguchi},
  {Okamura}, {Ly}, {Iye}, {Furusawa}, {Shioya}, {Shibuya}, {Ishizaki}, \&
  {Toshikawa}}]{Kashikawa2011}
{Kashikawa} N. {et~al.}, 2011, \apj, 734, 119

\bibitem[{{Kubota} {et~al}\mbox{.}(2017){Kubota}, {Yoshiura}, {Takahashi},
  {Hasegawa}, {Yajima}, {Ouchi}, {Pindor}, \& {Webster}}]{Kubota2017}
{Kubota} K., {Yoshiura} S., {Takahashi} K., {Hasegawa} K., {Yajima} H., {Ouchi}
  M., {Pindor} B., {Webster} R.~L., 2017, ArXiv e-prints

\bibitem[{{Leitherer} {et~al}\mbox{.}(1999){Leitherer}, {Schaerer}, {Goldader},
  {Delgado}, {Robert}, {Kune}, {de Mello}, {Devost}, \&
  {Heckman}}]{leitherer1999}
{Leitherer} C. {et~al.}, 1999, \apjs, 123, 3

\bibitem[{{Ouchi} {et~al}\mbox{.}(2018){Ouchi}, {Harikane}, {Shibuya},
  {Shimasaku}, {Taniguchi}, {Konno}, {Kobayashi}, {Kajisawa}, {Nagao}, {Ono},
  {Inoue}, {Umemura}, {Mori}, {Hasegawa}, {Higuchi}, {Komiyama}, {Matsuda},
  {Nakajima}, {Saito}, \& {Wang}}]{Ouchi2018}
{Ouchi} M. {et~al.}, 2018, \pasj, 70, S13

\bibitem[{{Planck Collaboration} {et~al}\mbox{.}(2016){Planck Collaboration},
  {Adam}, {Aghanim}, {Ashdown}, {Aumont}, {Baccigalupi}, {Ballardini},
  {Banday}, {Barreiro}, {Bartolo}, {Basak}, {Battye}, {Benabed}, {Bernard},
  {Bersanelli}, {Bielewicz}, {Bock}, {Bonaldi}, {Bonavera}, {Bond}, {Borrill},
  {Bouchet}, {Boulanger}, {Bucher}, {Burigana}, {Calabrese}, {Cardoso},
  {Carron}, {Chiang}, {Colombo}, {Combet}, {Comis}, {Couchot}, {Coulais},
  {Crill}, {Curto}, {Cuttaia}, {Davis}, {de Bernardis}, {de Rosa}, {de Zotti},
  {Delabrouille}, {Di Valentino}, {Dickinson}, {Diego}, {Dor{\'e}}, {Douspis},
  {Ducout}, {Dupac}, {Elsner}, {En{\ss}lin}, {Eriksen}, {Falgarone}, {Fantaye},
  {Finelli}, {Forastieri}, {Frailis}, {Fraisse}, {Franceschi}, {Frolov},
  {Galeotta}, {Galli}, {Ganga}, {G{\'e}nova-Santos}, {Gerbino}, {Ghosh},
  {Gonz{\'a}lez-Nuevo}, {G{\'o}rski}, {Gruppuso}, {Gudmundsson}, {Hansen},
  {Helou}, {Henrot-Versill{\'e}}, {Herranz}, {Hivon}, {Huang}, {Ili{\'c}},
  {Jaffe}, {Jones}, {Keih{\"a}nen}, {Keskitalo}, {Kisner}, {Knox},
  {Krachmalnicoff}, {Kunz}, {Kurki-Suonio}, {Lagache}, {L{\"a}hteenm{\"a}ki},
  {Lamarre}, {Langer}, {Lasenby}, {Lattanzi}, {Lawrence}, {Le Jeune},
  {Levrier}, {Lewis}, {Liguori}, {Lilje}, {L{\'o}pez-Caniego}, {Ma},
  {Mac{\'{\i}}as-P{\'e}rez}, {Maggio}, {Mangilli}, {Maris}, {Martin},
  {Mart{\'{\i}}nez-Gonz{\'a}lez}, {Matarrese}, {Mauri}, {McEwen}, {Meinhold},
  {Melchiorri}, {Mennella}, {Migliaccio}, {Miville-Desch{\^e}nes}, {Molinari},
  {Moneti}, {Montier}, {Morgante}, {Moss}, {Naselsky}, {Natoli}, {Oxborrow},
  {Pagano}, {Paoletti}, {Partridge}, {Patanchon}, {Patrizii}, {Perdereau},
  {Perotto}, {Pettorino}, {Piacentini}, {Plaszczynski}, {Polastri}, {Polenta},
  {Puget}, {Rachen}, {Racine}, {Reinecke}, {Remazeilles}, {Renzi}, {Rocha},
  {Rossetti}, {Roudier}, {Rubi{\~n}o-Mart{\'{\i}}n}, {Ruiz-Granados},
  {Salvati}, {Sandri}, {Savelainen}, {Scott}, {Sirri}, {Sunyaev}, {Suur-Uski},
  {Tauber}, {Tenti}, {Toffolatti}, {Tomasi}, {Tristram}, {Trombetti},
  {Valiviita}, {Van Tent}, {Vielva}, {Villa}, {Vittorio}, {Wandelt}, {Wehus},
  {White}, {Zacchei}, \& {Zonca}}]{Planck2016}
{Planck Collaboration} {et~al.}, 2016, \aap, 596, A108

\bibitem[{{Salpeter}(1955)}]{salpeter1955}
{Salpeter} E.~E., 1955, \apj, 121, 161

\bibitem[{{Sobacchi} {et~al}\mbox{.}(2016){Sobacchi}, {Mesinger}, \&
  {Greig}}]{Sobacchi2016}
{Sobacchi} E., {Mesinger} A., {Greig} B., 2016, \mnras, 459, 2741

\bibitem[{{Springel} \& {Hernquist}(2003)}]{springel2003}
{Springel} V., {Hernquist} L., 2003, \mnras, 339, 289

\bibitem[{{Vrbanec} {et~al}\mbox{.}(2016){Vrbanec}, {Ciardi}, {Jeli{\'c}},
  {Jensen}, {Zaroubi}, {Fernandez}, {Ghosh}, {Iliev}, {Kakiichi}, {Koopmans},
  \& {Mellema}}]{Vrbanec2016}
{Vrbanec} D. {et~al.}, 2016, \mnras, 457, 666

\bibitem[{{Wiersma} {et~al}\mbox{.}(2013){Wiersma}, {Ciardi}, {Thomas},
  {Harker}, {Zaroubi}, {Bernardi}, {Brentjens}, {de Bruyn}, {Daiboo}, {Jelic},
  {Kazemi}, {Koopmans}, {Labropoulos}, {Martinez}, {Mellema}, {Offringa},
  {Pandey}, {Schaye}, {Veligatla}, {Vedantham}, \& {Yatawatta}}]{Wiersma2013}
{Wiersma} R.~P.~C. {et~al.}, 2013, \mnras, 432, 2615

\bibitem[{{Wyithe} \& {Loeb}(2007)}]{Wyithe2007}
{Wyithe} J.~S.~B., {Loeb} A., 2007, \mnras, 375, 1034

\end{thebibliography}

\appendix

\end{document}